\documentclass[letterpaper,11pt]{article}

\usepackage[scale=0.75]{geometry}
\usepackage{amsmath}
\usepackage{amsthm}
\usepackage{amssymb}
\usepackage{mathtools}
\usepackage{times}
\usepackage[T1]{fontenc}
\usepackage{enumerate}
\usepackage{ifthen}

\DeclarePairedDelimiter{\abs}{\lvert}{\rvert}

\DeclarePairedDelimiter{\floor}{\lfloor}{\rfloor}
\DeclarePairedDelimiter{\ceil}{\lceil}{\rceil}

\DeclarePairedDelimiter{\ket}{\lvert}{\rangle}

\newcommand{\CC}{\mathbb{C}}
\newcommand{\NN}{\mathbb{N}}

\DeclareMathOperator{\rmspan}{span}
\DeclareMathOperator{\wsize}{wsize}

\newcommand{\calH}{\mathcal{H}}

\newcommand{\Nei}{\mathrm{N}}

\theoremstyle{plain}
\newtheorem{theorem}{Theorem}
\newtheorem{lemma}[theorem]{Lemma}
\newtheorem{corollary}[theorem]{Corollary}

\theoremstyle{definition}
\newtheorem{definition}{Definition}

\newcommand{\m}[2][]{\ifthenelse{\equal{}{#1}}%
 { \begin{displaymath} #2 \end{displaymath} }%
 { \begin{equation} \label{#1} #2 \end{equation} }%
}

\newcommand{\mal}[2][]{\m[#1]{\begin{aligned} #2 \end{aligned}}}

\newcommand{\PR}[2][]{\mathop{\mathbf{Pr}}_{#1}{\left[{#2}\right]}}

\newcommand{\E}[2][]{\mathop{\mathbf{E}}_{#1}{\left[{#2}\right]}}

\newcommand{\chs}{\genfrac(){0cm}{}}

\newcommand{\tm}{\cdot}

\title{A quantum query algorithm for the graph collision problem}

\author{
  \textbf{Dmitry Gavinsky}\\
  {\small NEC Laboratories America, Inc.}
 \and
  \textbf{Tsuyoshi Ito}\\
  {\small NEC Laboratories America, Inc.}
}

\date{}

\begin{document}
\maketitle

\begin{abstract}
We construct a new quantum algorithm for the graph collision problem;
that is, the problem of deciding whether the set of marked vertices contains a pair of adjacent vertices in a known graph~$G$.
The query complexity of our algorithm is $O(\sqrt{n}+\sqrt{\alpha^*(G)})$, where~$n$ is the number of vertices and~$\alpha^*(G)$ is the maximum total degree of the vertices in an independent set of $G$.
Notably, if~$G$ is a random graph where every edge is present with a fixed probability independently of other edges, then our algorithm requires~$O(\sqrt{n\log n})$ queries on most graphs, which is optimal up to the~$\sqrt{\log n}$ factor on most graphs.
\end{abstract}

\section{Introduction}

The quantum query complexity of a function is a natural counterpart of its classical query complexity; namely, it is the number of quantum oracle calls that an algorithm has to make in order to compute its value.
Grover's search algorithm~\cite{G96_A_F} gave the first example of a function whose quantum query complexity is significantly smaller than classical: computing the value of the OR function $\bigvee_{i=1}^nx_i$ requires $\Omega(n)$ classical queries, but only $O(\sqrt n)$ quantum queries.

It has been shown by Bennett et al.~\cite{BenBerBraVaz97SICOMP} that Grover's algorithm is optimal for computing the OR function.
Later, Beals et al.~\cite{BBCMW01_Qua_Lo} proved that no total function can have the gap between its quantum and classical query complexities larger than polynomial.

Another important problem where a quantum query algorithm can be much faster than a classical one is the element distinctness problem, where $n$ elements are ``colored'' by an oracle and the goal is to decide whether there are at least two elements of the same color.
In 2001 Buhrman et al.~\cite{BDHHM05_Qu_A} constructed an algorithm that required $O(n^{3/4})$ quantum queries, later a lower bound of $\Omega(n^{2/3})$ was shown by Aaronson and Shi~\cite{AS04_Qu_Lo}.
Finally, in 2003 Ambainis~\cite{Ambainis07SICOMP} gave a new algorithm that had query complexity $O(n^{2/3})$, thus matching the lower bound.

The graph collision problem was first considered by Magniez et al.~\cite{MagSanSze07SICOMP}, where it was shown to have quantum query complexity $O(n^{2/3})$.
The algorithm used in~\cite{MagSanSze07SICOMP} can be viewed as a natural adaptation of Ambainis' algorithm for the element distinctness problem.
On the other hand, the lower bound techniques used in~\cite{AS04_Qu_Lo} don't seem to be applicable to the graph collision problem, and the actual quantum query complexity of it is still an open question.

\subsection{Our results and techniques}

We present a new quantum algorithm for the graph collision problem.
The complexity of our algorithm depends on the properties of the given graph~$G$.
Throughout the paper, the quantum query complexity of a decision problem refers to that with a constant two-sided error.

\begin{theorem} \label{theorem:main}
For a graph~$G$ on~$n$ vertices,
the quantum query complexity of the graph collision problem on~$G$ is~$O(\sqrt{n}+\sqrt{\alpha^*(G)})$,
where $\alpha^*(G)$ is the maximum total degree of the vertices in an independent set of $G$.
\end{theorem}

Notably, this implies that the graph collision problem requires only~$\tilde{O}(\sqrt{n})$ quantum queries for most graphs in the following sense.
Let~$\mu_{n,p}$ be the distribution corresponding to choosing a graph on $n$ vertices, where every edge is present with probability~$p$ independently of other edges.

\begin{corollary} \label{corollary:main}
For arbitrary function $p\colon\NN\to[0,1]$, let $G\sim\mu_{n,p(n)}$.
Then the (worst-case) quantum query complexity of the graph collision problem on~$G$ is almost always\footnote
{Cf.\ Theorem~\ref{theorem:precor} for the corresponding quantitative statement.}
$O(\sqrt{n\log n})$.
\end{corollary}

The above result is optimal up to the $\sqrt{\log n}$ factor for most random graphs, as computing the OR of~$n$ variables can be reduced to solving the graph collision problem on any graph~$G$ that contains~$\Omega(n)$ non-isolated vertices.

Our algorithm for Theorem~\ref{theorem:main} works as follows.
As a preprocessing, we estimate the sum of the degrees of the vertices in $S$.
If this sum is much larger than $\max\{\alpha^*(G),n\}$, then we answer ``$S$ is not an independent set'' and halt.
This requires $O(\sqrt{n})$ queries, due to the approximate counting algorithm
by Brassard, H\o yer, and Tapp~\cite{BraHoyTap98ICALP}.
To handle the remaining (main) case, we construct a span program with witness size~$O(\sqrt{n}+\sqrt{\alpha^*(G)})$.
It was shown by Reichardt~\cite{Reichardt09FOCS,Reichardt11SODA}
that the quantum query complexity of a promise decision problem
is at most a constant factor away from the witness size of a span program computing it.

\subsection{Related work}

Magniez, Santha, and Szegedy~\cite{MagSanSze07SICOMP}
introduced the graph collision problem
and gave a quantum algorithm with~$O(n^{2/3})$ queries.
They used it as a subroutine used in their~$O(n^{13/10})$-query algorithm for the triangle finding problem.
This~$O(n^{2/3})$ is the best known upper bound
on the quantum query complexity
of the graph collision problem.
The best known lower bound for the graph collision problem
is~$\Omega(\sqrt{n})$, which follows easily from the lower bound
for the search problem~\cite{BenBerBraVaz97SICOMP}.
Jeffery, Kothari, and Magniez~\cite{JefKotMag-1112.5855v2}
recently gave a quantum algorithm
for the graph collision problem on a bipartite graph
which is useful when the given bipartite graph is close to the complete bipartite graph:
the query complexity of their algorithm is~$\tilde{O}(\sqrt{n}+\sqrt{m})$,
where~$m$ is the number of missing edges compared to the complete bipartite graph.

Improving the query complexity of the graph collision problem
has important consequences.
First, improving it is likely to give a better algorithm for the triangle finding problem
by applying the same technique as the one used in Ref.~\cite{MagSanSze07SICOMP}.
Second, the graph collision problem is equivalent to the evaluation of a 2-DNF formula,
and the techniques used in the graph collision problem
may be also applicable to the more general~$k$-DNF evaluation.

Our algorithm for the main case of the graph collision problem,
including its use of span programs,
is inspired by the recent result by Belovs and Lee~\cite{BelLee-1108.3022v1}.

\section{Preliminaries}

We will consider the quantum query complexity of the following problem.

\begin{definition}[Graph collision problem]
Let $G=(V,E)$ be a graph.
The \emph{graph collision problem on~$G$} asks, given oracle access to a string $x\in\{0,1\}^V$,
whether there exists an edge~$(i,j)\in E$ such that~$x_i=x_j=1$.
\end{definition}

Note that graph~$G$ is given explicitly to the algorithm, and the only part of the input which needs to be queried is the string~$x\in\{0,1\}^V$.
We call a vertex~$i$ \emph{marked} if~$x_i=1$.
Note that the graph collision problem is equivalent to deciding whether the marked vertices form an independent set in~$G$, with the answers ``yes'' and ``no'' swapped.

In the rest of the paper, we let~$V=[n]$, where~$[n]$ denotes the set~$\{1,\dots,n\}$.
For a graph~$G=(V,E)$ and a set~$S\subseteq V$ of vertices,
we denote by~$\deg(S)$ the sum of degrees of vertices in~$S$.
For any graph~$G$, we denote by~$\alpha^*(G)$ the maximum total degree of the vertices in an independent set of $G$; that is,
\[
  \alpha^*(G) = \max\{\deg(S) \colon \text{$S$ is an independent set in~$G$}\}.
\]

We will use the following form of Chernoff bound, as stated by Drukh and Mansour~\cite{DM05_Conc}.

\begin{lemma}[Chernoff bound] \label{l:Che}
Let~$X_1,\dots,X_n$ be mutually independent random variables taking values in~$[0,1]$, such that~$\E{X_i}=\mu$ for all~$i\in[n]$.
Then for any~$\lambda>1$,
\m{\PR{\sum_{i\in[n]}X_i\ge\lambda n\mu}\le
 \exp\left(-\frac{n(\lambda-1)^2\mu}{\lambda+1}\right)\le
 \exp\bigl((3-\lambda)n\mu\bigr).}
\end{lemma}

All logarithms in this paper are natural.

\subsection{Span programs}

\emph{Span program} is a linear-algebraic model of computation
introduced by Karchmer and Wigderson~\cite{KarWig93CoCo}
to study the computational power
of counting in branching programs and space-bounded computation.
In our context, the relevant complexity measure is its \emph{witness size}
introduced by Reichardt and \v{S}palek~\cite{ReiSpa08STOC}.
We use a formulation closer to that used by Reichardt~\cite{Reichardt09FOCS}.

\begin{definition}[Span program]
  \label{definition:span}
  A \emph{span program}~$P=(\calH,\ket{t};V_{10},V_{11},\dots,V_{n0},V_{n1})$
  with~$n$-bit input
  is defined by a finite-dimensional Hilbert space~$\calH$ over~$\CC$,
  a vector~$\ket{t}\in\calH$, and
  a finite set~$V_{jb}\subseteq\calH$ for each~$j\in[n]$ and each~$b\in\{0,1\}$.
  This span program is said to \emph{compute} a function~$f\colon D\to\{0,1\}$,
  where~$D\subseteq\{0,1\}^n$,
  when for~$x\in D$, we have~$f(x)=1$ if and only if
  $\ket{t}$ lies in the subspace of~$\calH$ spanned by~$\bigcup_{j\in[n]}V_{jx_j}$.
  The vector~$\ket{t}$ is called the \emph{target vector} of this span program~$P$.
\end{definition}

\begin{definition}[Witness size of a span program]
  Let~$P$ be a span program as in Definition~\ref{definition:span}.
  \begin{itemize}
  \item
    For an input~$x\in f^{-1}(1)$, a \emph{witness} for~$x$
    is an $n$-tuple of mappings~$w_1,\dots,w_n$,
    where~$w_j\colon V_{jx_j}\to\CC$,
    such that~$\ket{t}=\sum_{j\in[n]}\sum_{\ket{v}\in V_{jx_j}}w_j(\ket{v})\ket{v}$.
    The \emph{witness size} on input~$x\in f^{-1}(1)$,
    denoted by~$\wsize(P,x)$,
    is defined as
    \[
      \wsize(P,x)=\min_{(w_1,\dots,w_n):\text{witness for~$x$ in~$P$}}
        \sum_{j\in[n]}\sum_{\ket{v}\in V_{jx_j}}\abs{w_j(\ket{v})}^2.
    \]
  \item
    For an input~$x\in f^{-1}(0)$, a \emph{witness} for~$x$
    is a vector~$\ket{w'}\in\calH$ such that~$\langle t|w'\rangle=1$
    and~$\langle v|w'\rangle=0$ for every~$\ket{v}\in\bigcup_{j\in[n]}V_{jx_j}$.
    This time, the \emph{witness size} on input~$x\in f^{-1}(0)$,
    again denoted by~$\wsize(P,x)$,
    is defined as
    \[
      \wsize(P,x)=\min_{\ket{w'}:\text{witness for~$x$ in~$P$}}
        \sum_{j\in[n],b\in\{0,1\}}\sum_{\ket{v}\in V_{jb}}\abs{\langle v|w'\rangle}^2.
    \]
  \item
    The \emph{witness size} of this span program
    is~$\wsize(P)=\max_{x\in D}\wsize(P,x)$.
  \item
    Finally, we denote by~$\wsize(f)$ the minimum witness size of a span program
    which computes~$f$.
  \end{itemize}
\end{definition}

For a function~$f\colon D\to\{0,1\}$, where~$D\subseteq\{0,1\}^n$,
we denote by~$Q(f)$ the quantum query complexity of~$f$
with two-sided error probability at most~$1/3$.
As is well known, changing the error probability to other constants less than~$1/2$
affects the query complexity only within a constant factor.

\begin{theorem}[Reichardt~\cite{Reichardt09FOCS,Reichardt11SODA}]
    \label{theorem:reichardt}
  Let~$f\colon D\to\{0,1\}$, where~$D\subseteq\{0,1\}^n$.
  Then~$Q(f)$ and~$\wsize(f)$ coincide up to a constant factor.
  That is, there exists a constant~$c>1$ which does not depend on~$n$ or~$f$
  such that~$(1/c)\wsize(f)\le Q(f)\le c\cdot\wsize(f)$.
\end{theorem}

\begin{proof}
  Ref.~\cite{Reichardt09FOCS} showed that~$\wsize(f)$
  is equal to the general adversary bound for~$f$,
  and Ref.~\cite{Reichardt11SODA} showed that
  the general adversary bound for~$f$ and~$Q(f)$ coincide
  up to a constant factor.
\end{proof}

\subsection{Quantum algorithm for approximate counting}

To detect the case where marked vertices have too many edges,
we will use the following result
by Brassard, H\o yer, and Tapp~\cite[Theorem~5]{BraHoyTap98ICALP}.

\begin{theorem}[Approximate counting~\cite{BraHoyTap98ICALP}]
    \label{theorem:counting}
  There exists a quantum algorithm which,
  given integers~$N\ge1$ (domain size) and~$P\ge4$ (precision)
  and oracle access to a function~$F\colon[N]\to\{0,1\}$
  satisfying~$t=\abs{F^{-1}(1)}\le N/2$,
  makes~$P$ queries to the oracle and outputs an integer~$\tilde{t}$,
  such that
  \[
    \abs{t-\tilde{t}}<\frac{2\pi}{P}\sqrt{tN}+\frac{\pi^2}{P^2}N
  \]
  with probability at least~$8/\pi^2$.
\end{theorem}

We can remove the assumption that~$t\le N/2$ by doubling the size of the domain,
and we can reduce the error probability to an arbitrarily small constant
by repeating the algorithm constantly many times and taking the majority vote:

\begin{corollary} \label{corollary:counting}
  Let~$\varepsilon>0$ be a constant.
  Then there exists a quantum algorithm which,
  given integers~$N\ge1$ (domain size) and~$P\ge4$ (precision)
  and oracle access to a function~$F\colon[N]\to\{0,1\}$,
  makes~$O(P)$ queries to the oracle
  and outputs an integer~$\tilde{t}$ satisfying the following.
  Let~$t=\abs{F^{-1}(1)}$.
  Then it holds that
  \[
    \abs{t-\tilde{t}}<\frac{2\sqrt2\,\pi}{P}\sqrt{tN}+\frac{2\pi^2}{P^2}N
  \]
  with probability at least~$1-\varepsilon$.
  (The constant factor hidden in the~$O$-notation of the number of queries
   depends only on~$\varepsilon$ and not on~$N$, $P$, or~$F$.)
\end{corollary}

\section{Algorithm for the main case}

The following lemma is useful
in the case where not too many edges are incident to marked vertices.
In the next section, we will use it with~$k=2\max\{n,\alpha^*(G)\}$
to prove Theorem~\ref{theorem:main}.

\begin{lemma} \label{lemma:main-case}
  Let~$G$ be a graph on~$n$ vertices, and let~$k\in\NN$.
  Consider the special case of the graph collision problem on~$G$
  where it is promised that the set~$S$ of marked vertices satisfies~$\deg(S)\le k$.
  \begin{enumerate}[(i)]
  \item
    There exists a span program for this promise problem
    whose witness size is at most~$\sqrt{2(n+k)}$.
  \item
    There exists a quantum algorithm for this promise problem
    with two-sided error probability at most~$1/6$
    whose query complexity is~$O(\sqrt{n}+\sqrt{k})$.
  \end{enumerate}
\end{lemma}

\begin{proof}
  Item~(ii) follows immediately from item~(i)
  and Theorem~\ref{theorem:reichardt}.
  In the rest of the proof, we will prove item~(i)
  by constructing a span program explicitly.

  Let~$\calH=\CC^{\{0,1\}^n}$, and let
  \[
    \ket{t}=\gamma\sum_{z\in\{0,1\}^n}\ket{z},
    \qquad
    \gamma=\biggl(\frac{n+k}{2}\biggr)^{1/4}.
  \]
  For~$j\in[n]$ and~$b\in\{0,1\}$,
  let~$\ket{s_{jb}}=\sum_{z\in\{0,1\}^n:z_j=b}\ket{z}$.
  Let~$V_{j0}=\varnothing$
  and~$V_{j1}=\{\ket{s_{j0}}\}\cup\{\ket{s_{i1}}\colon i\in\Nei(j)\}$,
  where~$\Nei(j)$ is the set of neighbors of vertex~$j$ in graph~$G$.
  Define a span program~$P$
  as~$P=(\calH,\ket{t};V_{10},V_{11},\dots,V_{n0},V_{n1})$.

  It is easy to see that~$P$ computes the promise problem stated in the lemma.
  Indeed, if~$x\in f^{-1}(1)$,
  then there exists an edge~$ij\in E$ such that~$x_i=x_j=1$.
  Therefore, $\ket{s_{j0}}\in V_{j1}$ and~$\ket{s_{j1}}\in V_{i1}$,
  which implies that~%
  $\ket{t}=\gamma\ket{s_{j0}}+\gamma\ket{s_{j1}}
   \in\rmspan(V_{j1}\cup V_{i1})$.
  On the other hand, if~$x\in f^{-1}(0)$,
  then~$\ket{w'}=\ket{x}/\gamma$ is a witness for~$x$.
  Indeed, $\langle t|w'\rangle=1$,
  $\langle s_{j0}|w'\rangle=0$ if~$x_j=1$,
  and~$\langle s_{i1}|w'\rangle=0$
  if~$x_j=1$ and~$i\in\Nei(j)$.

  From these witnesses, the witness size of~$P$ can be bounded easily.
  If~$x\in f^{-1}(1)$,
  then the witness stated above
  shows that the witness size for~$x$ is at most~$2\gamma^2=\sqrt{2(n+k)}$.
  If~$x\in f^{-1}(0)$,
  then the witness stated above
  shows that the witness size for~$x$ is at most~$(n+k)/\gamma^2=\sqrt{2(n+k)}$.
\end{proof}

\section{Preprocessing and overall algorithm}

In this section, we will prove Theorem~\ref{theorem:main}.

Consider the following quantum algorithm.
\begin{enumerate}
\item
  Compute~$\alpha^*(G)$.
  Let~$s=\max\{\alpha^*(G),n\}$.
  (Because this step does not use any queries to~$x$,
   how~$\alpha^*(G)$ is computed does not matter
   as long as the query complexity is concerned.)
\item
  (Preprocessing.)
  Estimate the number~$t$ of pairs~$(i,j)\in[n]^2$ such that~$ij\in E$ and~$x_i=1$
  by running the approximate counting algorithm in Corollary~\ref{corollary:counting}
  with error probability~$\varepsilon=1/6$
  and precision parameter~$P=\max\{4,\ceil{7\pi\sqrt{n}}\}$.
  If the result~$\tilde{t}$ of counting satisfies~$\tilde{t}>3s/2$,
  then answer ``yes'' and halt.
  Otherwise, proceed to the next step.
\item
  (Main case.)
  Run the algorithm in Lemma~\ref{lemma:main-case}~(ii)
  using error probability~$1/6$ and parameter~$k=2s$,
  and answer ``yes'' or ``no'' accordingly.
\end{enumerate}

\paragraph{Query complexity.}
Step~1 does not make any queries to~$x$.
Step~2 makes~$O(P)=O(\sqrt{n})$ queries.
Step~3 makes~%
$O(\sqrt{n}+\sqrt{2s})
 =O(\sqrt{n}+\sqrt{\alpha^*(G)})$ queries.
Therefore, the query complexity of the whole algorithm
is~$O(\sqrt{n}+\sqrt{\alpha^*(G)})$.

\paragraph{Correctness.}
In the rest of this section, we will show that this algorithm reports an incorrect answer
with probability at most~$1/3$ by considering the following three cases:
(a) the correct answer is ``yes'' and~$t\le2s$,
(b) the correct answer is ``yes'' and~$t>2s$,
(c) the correct answer is ``no.''

If the correct answer is ``yes'' and~$t\le2s$,
the only step where the algorithm reports an incorrect answer is step~3,
and the promise of the algorithm in Lemma~\ref{lemma:main-case}~(ii) is satisfied.
Therefore, the error probability is at most~$1/6$.

If the correct answer is ``yes'' and~$t>2s$,
then with probability at least~$5/6$, $\tilde{t}$ satisfies that
\begin{align*}
  \tilde{t}
  &\ge
  t-\abs{t-\tilde{t}}
  \\
  &>
  t-\frac{2\sqrt2\,\pi}{P}\sqrt{tn^2}-\frac{2\pi^2}{P^2}n^2
  \displaybreak[0] \\
  &\ge
  t-\frac{2\sqrt2}{7}\sqrt{tn}-\frac{2}{49}n
  \displaybreak[0] \\
  &\ge
  \sqrt{t}\Bigl(\sqrt{t}-\frac{2\sqrt2}{7}\sqrt{n}\Bigr)-\frac{2}{49}n
  \\
  &\ge
  \sqrt{2s}\Bigl(\sqrt{2s}-\frac{2\sqrt2}{7}\sqrt{s}\Bigr)-\frac{2}{49}s
  >
  \frac32 s.
\end{align*}
Therefore, the algorithm reports ``yes'' in step~2 alone with probability at least~$5/6$.
In this case, the promise of the algorithm in Lemma~\ref{lemma:main-case}~(ii) 
is not satisfied, but it does not matter what step~3 reports.

Finally, consider the case where the correct answer is ``no.''
In this case, both steps~2 and~3 can report an incorrect answer,
and we will bound each of these probability by~$1/6$.
The correct answer being ``no''
means that the set of marked vertices is an independent set of~$G$,
and therefore it holds that~$t\le\alpha^*(G)\le s$
by the definitions of~$\alpha^*(G)$ and~$s$.
This implies that with probability at least~$5/6$, $\tilde{t}$ satisfies that
\begin{align*}
  \tilde{t}
  &\le
  t+\abs{t-\tilde{t}}
  \\
  &<
  t+\frac{2\sqrt2\,\pi}{P}\sqrt{tn^2}+\frac{2\pi^2}{P^2}n^2
  \displaybreak[0] \\
  &\le
  t+\frac{2\sqrt2}{7}\sqrt{tn}+\frac{2}{49}n
  \\
  &\le
  s+\frac{2\sqrt2}{7}\sqrt{s\cdot s}+\frac{2}{49}s
  <
  \frac32 s.
\end{align*}
Therefore, step~2 reports in an incorrect answer with probability at most~$1/6$.
Moreover, because the promise of the algorithm
in Lemma~\ref{lemma:main-case}~(ii) is satisfied,
step~3 reports an incorrect answer with probability at most~$1/6$.
By union bound, the overall error probability is at most~$1/6+1/6=1/3$.

\section{The case of random graphs}

In this section we analyze the query complexity of the graph collision problem defined over random graphs.
Recall that we denote by $\mu_{n,p}$ the distribution of random graphs on $n$ vertices, where every edge is present with probability $p$, independently of other edges.

We need the following combinatorial lemma.

\begin{lemma} \label{l:al*G_ra}
For arbitrary $p\in(0,1]$ and $t\ge 40n\log n$,
\m{\PR[G\sim\mu_{n,p}]{\alpha^*(G)\ge t}
 \le n^{-14n} + 2\exp\left(\frac{-t^2}{200n^2p}\right).}
\end{lemma}

\begin{proof}
Assume $G\sim\mu_{n,p}$.
For any $t\in\NN$, let $Y_t$ be the expected number of independent sets $S\subseteq[n]$ that satisfy $\deg(S)\ge t$.
Clearly,
\m{\PR{\alpha^*(G)\ge t}\le\E{Y_t},}
and therefore we want an upper bound on $\E{Y_t}$.

For any $i\ge2$ it holds that
\m{\chs ni\tm(1-p)^{\chs i2}
 \le\exp\left(i\log n-\frac{pi^2}4\right).}
Let $x_0\in[n]$, to be fixed later.
Then
\mal{\E{Y_t}
 &\le\sum_{i=1}^{x_0}\chs ni\tm(1-p)^{\chs i2}\tm
  \PR{\deg(S)\ge t~:~|S|=i}\\
 &~~+\sum_{i=x_0+1}^n\chs ni\tm(1-p)^{\chs i2}\\
 &\le \sum_{i=1}^{x_0}\exp(i\log n + 3nip-t)
  +\sum_{i=x_0+1}^n\exp\left(i\log n-\frac{pi^2}4\right),}
where the last inequality follows from Lemma~\ref{l:Che}.

Fix $x_0=\min\bigl\{\floor[\big]{\frac t{5np}},n\bigr\}$.
Then, noting~$t\ge 40n\log n$, it holds that
\m{\sum_{i=x_0+1}^n\exp\left(i\log n-\frac{pi^2}4\right)
 \le2\exp\left(-\frac{px_0^2}8\right),}
and we continue:
\mal{\E{Y_t} &\le\exp(\log x_0+x_0\log n+3nx_0p-t)
  +2\exp\left(-\frac{px_0^2}8\right)\\
 &\le\exp\left(-\frac{7t}{20}\right)
  +2\exp\left(-\frac{t^2}{200n^2p}\right).}
The result follows.
\end{proof}

The theorem below and Corollary~\ref{corollary:main} follow immediately from Theorem~\ref{theorem:main} and Lemma~\ref{l:al*G_ra}.

\begin{theorem} \label{theorem:precor}
There exists a universal constant $C$ such that for any $p\in(0,1]$, $n\in\NN$ and $t\ge 40n\log n$ the following holds.
For $G\sim\mu_{n,p}$, the probability that the (worst-case) quantum query complexity of the graph collision problem on~$G$ is greater than $C(\sqrt n+\sqrt t)$ is at most $n^{-14n} + 2\exp\bigl(\frac{-t^2}{200n^2p}\bigr)$.
\end{theorem}

\section{Concluding remarks}

We gave a quantum algorithm for the graph collision problem
on graph~$G$ on~$n$ vertices
whose query complexity is bounded as~$O(\sqrt{n}+\sqrt{\alpha^*(G)})$
in terms of the maximum sum of degrees of the vertices in an independent set of~$G$.
We used this to show that the graph collision problem
has quantum query complexity~$\tilde{O}(\sqrt{n})$ for almost all graphs
if a graph is chosen at random
so that each edge is present with a fixed probability independently of other edges.

We conclude by stating a few open problems.
Clearly improving the algorithm so that its query complexity becomes~$\tilde{O}(\sqrt{n})$ for \emph{all} graphs is an important open problem.
As another direction, the graph collision problem can be defined also for hypergraphs,
and it is used in an algorithm for the subgraph finding problem~\cite{MagSanSze07SICOMP}, a natural generalization of the triangle finding problem.
Extending the present algorithm to the case of hypergraphs is another open problem.

\subsection*{Acknowledgments}

Dmitry Gavinsky is grateful to Ryan O'Donnell, Rocco Servedio, Srikanth Srinivasan and Li-Yang Tan for helpful discussions.
The authors acknowledge support by ARO/NSA under grant W911NF-09-1-0569.

\bibliography{graph-collision}

\end{document}